\documentclass[11pt,a4paper]{article}
\usepackage[english]{babel}
\usepackage{amsmath,amssymb,amsfonts}
\usepackage{booktabs}
\usepackage{geometry}
\usepackage{graphicx}
\usepackage{bm}
\usepackage{float}
\begin{document}

\begin{center}
{\Large\bf Linearity, Nonlinearity and Universality of Regge Trajectories in Light Mesons: a Statistical Approach}
\end{center}

\begin{center}
S.S. Afonin\footnote{E-mail: \texttt{s.afonin@spbu.ru}}
\end{center}

\begin{center}
  {\small Saint Petersburg State University, 
  St. Petersburg 199034 Russia}\\
  \vspace*{0.15cm}
  {\small NRC "Kurchatov Institute" -- PNPI, Gatchina 188300 Russia
  }
\end{center}

\begin{abstract}
The masses of light non-strange mesons can be parameterized as a function of the radial quantum number $n$ and the orbital angular momentum $L$. We perform a comprehensive statistical and phenomenological comparative analysis of competing Regge-like formulas evaluated against two experimental datasets: a benchmark sample of 27 well-established states from the 2024 Particle Data Group (PDG) data, and an expanded sample of 85 states compiled within a recently proposed $(L,n)$-classification. Two linear Regge models for $M^2(L,n)$ are tested: one featuring distinct slopes for $L$ and $n$, $M^2(L,n)\propto an+bL$ (Model~1), and another assuming a universal slope, $M^2(L,n)\propto a(n+L)$ (Model~2). Model selection is conducted quantitatively using the Residual Sum of Squares (RSS), the adjusted RSS, the Akaike Information Criterion (AIC), and the Bayesian Information Criterion (BIC). Within the 27 benchmark states, Model~2 is statistically disfavored. We argue, however, that this discrepancy is driven by the specific behavior of $S$-wave ($L=0$) states. Upon their exclusion, the difference in performance between the two models becomes statistically insignificant, with information criteria selecting Model~2 as the more parsimonious description. Furthermore, we demonstrate that the masses of $S$-wave states can be successfully accommodated by a physically motivated, nonlinear $L$-dependent correction. A parallel analysis of the extended 85-state dataset, where the relative contribution of $S$-wave states is significantly smaller, reveals that Model~2 is statistically preferred, thereby restoring the Coulomb-like degeneracy in the Regge spectrum under consideration.
\end{abstract}

\section{Introduction}

The physics of confinement in QCD is clearly manifested in the highly organized structure of the spectrum of light hadrons.
The mass spectrum of light mesons composed of $u$ and $d$ quarks has long been served as a testing ground for models
of strong interactions in the non-perturbative regime.
As a direct derivation of these masses from the fundamental QCD Lagrangian remains hindered by the non-perturbative
strong coupling at low energies, hadron phenomenology relies heavily on effective models, dispersive methods, dualities, and other frameworks. Among these approaches, the concept of Regge trajectories --- wherein the hadron mass squared scales linearly with its quantum numbers --- remains a cornerstone of hadron phenomenology~\cite{50years}. Originally discovered within the framework of $S$-matrix theory~\cite{Collins:1977}, the linear relationship between a meson's squared mass and its orbital angular momentum $L$ or radial excitation number $n$ has found strong theoretical justification in relativistic string and flux-tube models~\cite{Nambu:1974}.

In the valence quark approximation, a light non-strange meson ($u, d$ quark content) can be viewed as a system of two massless or nearly massless constituents connected by a color-confining flux tube with a constant string tension $\sigma$. Semiclassical quantization of a rotating relativistic string yields the famous Regge trajectory relation,
\begin{equation}
    M^2 = 2\pi\sigma L + c_0,
\end{equation}
where the orbital slope $b = 2\pi\sigma$ is universal for a given flavor sector~\cite{Collins:1977}. When extended to include radial pulsations of the string, the generalized linear ansatz typically reads (a recent discussion is given in~\cite{Afonin:2024egd}),
\begin{equation}
\label{eq:linear_gen}
    M^2 = an + bL + c,
\end{equation}
where $a$ represents the radial slope and $b$ denotes the orbital slope.

A long-standing debate in contemporary hadron physics centers on the structural relationship between these two slopes, known as the question of trajectory degeneracy. Some semiclassical hadron string considerations~\cite{Afonin:2024egd} and certain implementations of the AdS/QCD holographic approach~\cite{Karch:2006} imply a strict universality, enforcing $a = b$ (i.e., $M^2 \propto n + L$), which implies that a single string tension dictates both spatial rotations and radial breathing modes. However, global statistical analyzes of the experimental data accumulated by collaborations such as Crystal Barrel at CERN~\cite{Anisovich:2000, Bugg:2004}  suggest that the radial trajectories may possess a significantly steeper slope than their angular-momentum counterparts ($a > b$), indicating a breakdown of naive universal degeneracy in the light meson spectrum~\cite{Masjuan:2012}.

Furthermore, while the asymptotic linearity of Eq.~(\ref{eq:linear_gen}) holds remarkably well for highly excited states ($n, L \gg 1$), various deviations
from this law are observed, particularly for the $S$-wave states ($L=0$). These distortions stem from short-range physical phenomena not captured by the long-range linear confinement potential. Specifically, one-gluon exchange  interactions introduce a Coulomb-like attractive behavior at small distances, while chiral symmetry breaking, spin-orbit couplings, and instanton-induced forces induce state-dependent mass shifts~\cite{Ebert:2009, Shifman:2008}. To account for these low-$L$ effects without destroying the successful asymptotic linearity, phenomenological models often introduce nonlinear corrections.

The aim of this paper is to perform a detailed, data-driven comparative analysis of these competing phenomenological frameworks. Utilizing a comprehensive benchmark dataset consisting of 27 well-established experimental mass states of light non-strange mesons~\cite{pdg} across different $(L,n)$ configurations, we perform unweighted least-squares regressions. We go beyond mere goodness-of-fit evaluations by extracting the full parameter sets alongside their 95\% confidence intervals. Crucially, to resolve the trade-off between model accuracy and parametric complexity, we employ information-theoretic model selection criteria, namely the Akaike Information Criterion (AIC) 
and the Bayesian Information Criterion (BIC). 
After this, we add unconfirmed states from PDG~\cite{pdg}, expanding the set to 85 states, and repeat our statistical analysis.
It is important to emphasize that any analysis of this kind depends on $(L,n)$-assignment of unconfirmed resonances. Our present analysis will be based on the modern $(L,n)$-classification constructed recently in~\cite{Afonin:2025dxk}.
This systematic approach aims to determine whether experimental data justify the inclusion of independent slopes or short-range nonlinear corrections, thereby shedding light on the underlying dynamics of quark confinement.

\section{Linear Regge and radial trajectories vs. experimental data on light non-strange mesons}

The spectrum of light non-strange mesons used in our analysis is shown in Table~1. The corresponding masses are given in Table~2. The pion is excluded due to its specific goldstone nature.

\begin{table}
\caption{
\small The modern $(L,n)$-classification of light non-strange mesons according to~\cite{Afonin:2025dxk}. The well-established states from PDG~\cite{pdg} are highlighted in bold.}
\vspace{-0.2cm}
\begin{center}
\resizebox{0.5\textwidth}{!}{
\begin{tabular}{|c|c|c|c|c|c|}
\hline
\begin{tabular}{c}
\begin{picture}(15,15)
\put(0,12){\line(1,-1){15}}
\put(-2,-3){$L$}
\put(10,7){$n$}
\end{picture}\\
\end{tabular}
& 0 & 1 & 2 & 3 & 4 \\
\hline
0
&
\begin{tabular}{c}
$\bm{\pi}$\\
---\\
$\bm{\rho}$\\
$\bm{\omega}$\\
\end{tabular}
&
\begin{tabular}{c}
$\bm{\pi(1300)}$\\
$\bm{\eta(1295)}$\\
$\bm{\rho(1450)}$ \\
$\bm{\omega(1420)}$ \\
\end{tabular}
&
\begin{tabular}{c}
$\bm{\pi(1800)}$\\
$\bm{\eta(1760)}$\\
$\rho(?)$\\
$\omega(?)$\\
\end{tabular}
&
\begin{tabular}{c}
$\pi(2070)$\\
$\eta(2010)$\\
$\rho(?)$ \\
$\omega(?)$\\
\end{tabular}
&
\begin{tabular}{c}
$\pi(2360)$\\
$\eta(2320)$\\
$\rho(?)$\\
$\omega(?)$ \\
\end{tabular}
\\
\hline
1
&
\begin{tabular}{c}
$\bm{a_0(1450)}$\\
$\bm{f_0(1370)}$\\
$\bm{a_1(1260)}$\\
$\bm{f_1(1285)}$\\
$\bm{b_1(1235)}$\\
$\bm{h_1(1170)}$\\
$\bm{a_2(1320)}$\\
$\bm{f_2(1270)}$\\
\end{tabular}
&
\begin{tabular}{c}
$a_0(1710)$\\
$f_0(1710)$\\
$\bm{a_1(1640)}$\\
$f_1(?)$\\
$b_1(?)$ \\
$h_1(?)$ \\
$\bm{a_2(1700)}$\\
$f_2(1750)$\\
\end{tabular}
&
\begin{tabular}{c}
$a_0(2020)$\\
$f_0(2020)$\\
$a_1(1930)$ \\
$f_1(1970)$\\
$b_1(1960)$\\
$h_1(1965)$\\
$a_2(2030)$ \\
$f_2(2000)$\\
\end{tabular}
&
\begin{tabular}{c}
$a_0(?)$\\
$f_0(2200)$\\
$a_1(2270)$ \\
$f_1(2310)$\\
$b_1(2240)$\\
$h_1(2215)$\\
$a_2(2175)$ \\ 
$f_2(2295)$\\
\end{tabular}
&\\
\hline
2
&
\begin{tabular}{c}
$\bm{\rho(1700)}$\\
$\bm{\omega(1650)}$\\
$\bm{\pi_2(1670)}$\\
$\bm{\eta_2(1645)}$\\
$\rho_2(?)$\\
$\omega_2(?)$\\
$\bm{\rho_3(1690)}$\\
$\bm{\omega_3(1670)}$\\
\end{tabular}
&
\begin{tabular}{c}
$\rho(2000)$\\
$\omega(1960)$\\
$\pi_2(2005)$\\
$\eta_2(2030)$\\
$\rho_2(1940)$\\
$\omega_2(1975)$\\
$\rho_3(1990)$\\
$\omega_3(1945)$\\
\end{tabular}
&
\begin{tabular}{c}
$\rho(2270)$\\
$\omega(2290)$ \\
$\pi_2(2285)$\\
$\eta_2(2250)$\\
$\rho_2(2225)$\\
$\omega_2(2195)$\\
$\rho_3(?)$ \\
$\omega_3(2285)$\\
\end{tabular}
&  &\\
\hline
3
&
\begin{tabular}{c}
$a_2(1990)$\\
$\bm{f_2(1950)}$\\
$a_3(2030)$\\
$f_3(2050)$\\
$b_3(2030)$\\
$h_3(2025)$\\
$\bm{a_4(1970)}$\\
$\bm{f_4(2050)}$\\
\end{tabular}
&
\begin{tabular}{c}
$a_2(2255)$ \\
$f_2(2240)$\\
$a_3(2275)$\\
$f_3(2300)$\\
$b_3(2245)$\\
$h_3(2275)$\\
$a_4(2255)$\\
$f_4(2300)$\\
\end{tabular}
&  &  &\\
\hline
4
&
\begin{tabular}{c}
$\rho_3(2250)$\\
$\omega_3(2255)$\\
$\pi_4(2250)$\\
$\eta_4(2330)$\\
$\rho_4(2230)$\\
$\omega_4(2250)$ \\
$\rho_5(2350)$\\
$\omega_5(2250)$\\
\end{tabular}
&  &  &  &\\
\hline
\end{tabular}
}
\end{center}
\end{table}

The physical principles for selection of states entering Table~1 and for definite $(L,n)$-assignment of selected states are discussed in detail in Ref.~\cite{Afonin:2025dxk}. We just make a couple of brief comments. First,
the letter "$\eta$" in the second row of Table~1, in reality, denotes the isosinglet pseudoscalar particle --- the given notation does not mean that the resonance under consideration has the same quark composition as the $\eta$-meson. The strange component gives the dominant contribution to the $\eta$-meson mass but the same cannot be claimed for the heavier resonances with the quantum numbers of $\eta$-meson according to the phenomenology of their production and decays.
Second, within the $(L,n)$-classification many highly excited meson resonances can be mapped by two ways. For instance, there are vector mesons with
$L=0$ ($S$-wave states) and $L=2$ ($D$-wave states). The general principle for resolving such uncertainties is that more reliable state should have
larger $L$ --- it is expected that the centrifugal barrier in the final state suppresses the decay (thus enhancing the probability of observation).

\begin{table}
\caption{
\small The experimental masses (in GeV) of states in Table 1~\cite{pdg}.}
\vspace{-0.2cm}
\begin{center}
\resizebox{0.5\textwidth}{!}{
\begin{tabular}{|c|c|c|c|c|c|}
\hline
\begin{tabular}{c}
\begin{picture}(15,15)
\put(0,12){\line(1,-1){15}}
\put(-2,-3){$L$}
\put(10,7){$n$}
\end{picture}\\
\end{tabular}
& 0 & 1 & 2 & 3 & 4 \\
\hline
0
&
\begin{tabular}{c}
---\\
---\\
$\mathbf{0.775}$\\
$\mathbf{0.783}$\\
\end{tabular}
&
\begin{tabular}{c}
$\mathbf{1.300\pm0.100}$\\
$\mathbf{1.294\pm0.004}$\\
$\mathbf{1.465\pm0.025}$\\
$\mathbf{1.410\pm0.060}$\\
\end{tabular}
&
\begin{tabular}{c}
$\mathbf{1.810\pm0.010}$\\
$\mathbf{1.751\pm0.015}$\\
---\\
---\\
\end{tabular}
&
\begin{tabular}{c}
$2.070\pm0.035$\\
$2.010\pm0.060$\\
---\\
---\\
\end{tabular}
&
\begin{tabular}{c}
$2.360\pm0.025$\\
$2.320\pm0.015$\\
---\\
---\\
\end{tabular}
\\
\hline
1
&
\begin{tabular}{c}
$\mathbf{1.439\pm0.034}$\\
$\mathbf{1.350\pm0.150}$\\
$\mathbf{1.230\pm0.040}$\\
$\mathbf{1.282\pm0.001}$\\
$\mathbf{1.230\pm0.003}$\\
$\mathbf{1.166\pm0.008}$\\
$\mathbf{1.318\pm0.001}$\\
$\mathbf{1.275\pm0.001}$\\
\end{tabular}
&
\begin{tabular}{c}
$1.713\pm0.019$\\
$1.733\pm0.008$\\
$\mathbf{1.655\pm0.016}$\\
---\\
---\\
---\\
$\mathbf{1.706\pm0.014}$\\
$1.755\pm0.010$\\
\end{tabular}
&
\begin{tabular}{c}
$2.025\pm0.030$\\
$1.982\pm0.057$\\
$1.930\pm0.070$\\
$1.971\pm0.015$\\
$1.960\pm0.035$\\
$1.965\pm0.045$\\
$2.030\pm0.020$\\
$2.001\pm0.010$\\
\end{tabular}
&
\begin{tabular}{c}
---\\
$2.187\pm0.014$\\
$2.270\pm0.055$\\
$2.310\pm0.060$\\
$2.240\pm0.035$\\
$2.215\pm0.040$\\
$2.175\pm0.040$\\
$2.293\pm0.013$\\
\end{tabular}
&\\
\hline
2
&
\begin{tabular}{c}
$\mathbf{1.720\pm0.020}$\\
$\mathbf{1.670\pm0.030}$\\
$\mathbf{1.671\pm0.002}$\\
$\mathbf{1.617\pm0.005}$\\
---\\
---\\
$\mathbf{1.689\pm0.002}$\\
$\mathbf{1.667\pm0.004}$\\
\end{tabular}
&
\begin{tabular}{c}
$2.000\pm0.030$\\
$1.960\pm0.025$\\
$1.963\pm0.027$\\
$2.030\pm0.020$\\
$1.940\pm0.040$\\
$1.975\pm0.020$\\
$1.982\pm0.014$\\
$1.945\pm0.020$\\
\end{tabular}
&
\begin{tabular}{c}
$2.265\pm0.040$\\
$2.290\pm0.020$\\
$2.285\pm0.045$\\
$2.248\pm0.020$\\
$2.225\pm0.035$\\
$2.195\pm0.030$\\
---\\
$2.285\pm0.060$\\
\end{tabular}
&  &\\
\hline
3
&
\begin{tabular}{c}
$2.050\pm0.050$\\
$\mathbf{1.936\pm0.012}$\\
$2.031\pm0.012$\\
$2.048\pm0.008$\\
$2.032\pm0.012$\\
$2.025\pm0.020$\\
$\mathbf{1.967\pm0.016}$\\
$\mathbf{2.018\pm0.011}$\\
\end{tabular}
&
\begin{tabular}{c}
$2.255\pm0.020$\\
$2.240\pm0.015$\\
$2.275\pm0.035$\\
$2.334\pm0.025$\\
$2.245\pm0.050$\\
$2.275\pm0.025$\\
$2.255\pm0.040$\\
$2.283\pm0.017$\\
\end{tabular}
&  &  &\\
\hline
4
&
\begin{tabular}{c}
$2.260\pm0.020$\\
$2.255\pm0.015$\\
$2.250\pm0.015$\\
$2.328\pm0.038$\\
$2.230\pm0.025$\\
$2.250\pm0.030$\\
$2.330\pm0.035$\\
$2.250\pm0.070$\\
\end{tabular}
&  &  &  &\\
\hline
\end{tabular}
}
\end{center}
\end{table}

We will fit the spectrum using three distinct Regge-like models, each carrying a different physical interpretation.
\begin{enumerate}
    \item \textbf{Model 1: Independent Slopes Model}
    \begin{equation}
        M^2 = an + bL + c.
    \end{equation}
    Here, $a$ represents the radial Regge trajectory slope, $b$ denotes the orbital Regge trajectory slope, and $c$ is the intercept corresponding to the ground-state mass squared ($n=0, L=0$). Physically, $a \neq b$ implies that exciting a meson radially requires a different amount of energy than increasing its orbital momentum.

    \item \textbf{Model 2: Universal Slope Model}
    \begin{equation}
        M^2 = a(n+L) + c.
    \end{equation}
    This formulation enforces a strict constraint $a = b$. In semiclassical flux-tube or string models, a universal slope directly implies a unique, undivided string tension $\sigma$ governing both radial and angular string stretching. This spectrum exhibits classical Coulomb-type degeneracy, i.e. dependence of excitation energy only on the principal quantum number $n+L+1$.

    \item \textbf{Model 3: Nonlinear Model with Universal Slope}
    \begin{equation}
        M^2 = a(n+L) + \frac{d}{L+1} + c.
    \end{equation}
    This ansatz re-introduces the universal slope $a$ and adds a nonlinear term governed by parameter $d$. The factor $d/(L+1)$ acts as a short-range or spin-orbit correction. As $L \to \infty$, this term vanishes, recovering the asymptotic linearity. A positive $d$ signifies an extra attractive contribution at low orbital momentum. The form of this correction is motivated by the first relativistic correction to the Coulomb spectrum of the Dirac particle in the Coulomb field~\cite{greiner}.
\end{enumerate}

\section{Global analysis}

As Table~2 shows, experimental uncertainties in meson mass determinations can vary by two orders of magnitude.
In this situation, the standard $\chi^2$ method (weighted least-squares) yields unreliable results and requires modification.
Experimental uncertainties in the masses of light mesons are not simply experimental errors as considered in statistics, but contain important information about the physical nature of a particular meson state: the extracted resonance mass is reaction-dependent (this is not an error but a manifestation of the underlying physics), and its value in the PDG is typically an average of the values obtained from different reactions.
The corresponding analysis using a modified $\chi^2$ method is planned for the future. In the present work,
we carry out a statistical analysis using the usual unweighted least-squares method for the central values.

First we consider subset containing $N=27$ experimental points corresponding to well established states in Table~2.
Then the whole set of $N=85$ experimental states from Table~2 is analyzed.
We implement multiple robust metrics to contrast the performance
of the models under consideration across these two datasets. The Residual Sum of
Squares (RSS) measures the amount of variance in a dataset that is not
explained by a regression model. The adjusted coefficient of
determination ($R^2_{\text{adj}}$) penalizes the standard
$R^2$ value based on the number of predictors. To resolve
the trade-off between model accuracy and parametric complexity,
we employ information-theoretic model selection criteria.
The Akaike Information Criterion ($\text{AIC} = 2k - N \ln(L)$, where $L$ is the
Maximum Likelihood function)
estimates information loss while penalizing the number of
free parameters ($k$) to prevent overfitting. The Bayesian
Information Criterion ($\text{BIC} = k \ln(N) - N \ln(L)$)
introduces a stricter complexity penalty based on the total
sample size ($N$). If a model is analyzed using the least-squares method and
the errors are normally distributed, the definitions of AIC and BIC can be simplified
through RSS,
\begin{equation}
\label{aic}
\text{AIC} = N \ln(\text{RSS}/N) + 2k + \text{const},
\end{equation}
\begin{equation}
\label{bic}
\text{BIC} = N \ln(\text{RSS}/N) + k \ln(N) + \text{const},
\end{equation}
where the usual choice for basic distribution constant in software is $\text{const}=N(\ln(2\pi)+1)$.
We have verified that the data is normally distributed with very good accuracy in all of our fits, so the
simplified definitions~\eqref{aic} and~\eqref{bic} can be used.
The obtained results are summarized in Table~\ref{tab:global_comparison}.

\begin{table}[htbp]
\centering
\caption{\small Extracted values of model parameters (in GeV$^2$) with $1\sigma$ errors, 95\% confidence intervals (CI),
and comprehensive goodness-of-fit metrics for the initial
($N=27$) and extended ($N=85$) meson datasets.}
\label{tab:global_comparison}
\vspace{0.3cm}
\resizebox{1\textwidth}{!}{
\begin{tabular}{lcccccc}
\toprule
\textbf{Dataset \& Model} & \textbf{Parameter} &
\textbf{Value $\pm$ Error / [95\% CI]} & \textbf{RSS} &
$R^2_{\text{adj}}$ & \textbf{AIC} & \textbf{BIC} \\
\midrule
\textbf{Initial Dataset ($N=27$)} & & & & & & \\
Model 1: $an + bL + c$ & $a$ & $1.275 \pm 0.066\ [1.140, 1.410]$ &
0.6961 & 0.9651 & -16.146 & -12.259 \\
                       & $b$ & $1.100 \pm 0.042\ [1.014, 1.186]$ &
                       & & & \\
                       & $c$ & $0.579 \pm 0.073\ [0.428, 0.729]$ &
                       & & & \\
\cmidrule{2-7}
Model 2: $a(n+L) + c$  & $a$ & $1.108 \pm 0.049\ [1.007, 1.209]$ &
1.0100 & 0.9513 & -8.095 & -5.503 \\
                       & $c$ & $0.631 \pm 0.084\ [0.457, 0.804]$ &
                       & & & \\
\cmidrule{2-7}
Model 3: $a(n+L) + \frac{d}{L+1} + c$ & $a$ & $1.215 \pm 0.052$ &
\textbf{0.6850} & \textbf{0.9678} & \textbf{-16.578} &
\textbf{-12.691} \\
                       & $d$ & $0.488 \pm 0.038$ & & & & \\
                       & $c$ & $0.183 \pm 0.041$ & & & & \\
\midrule
\textbf{Extended Dataset ($N=85$)} & & & & & & \\
Model 1: $an + bL + c$ & $a$ & $1.139 \pm 0.043\ [1.096, 1.182]$ &
2.7749 & 0.9795 & -284.87 & -277.55 \\
                       & $b$ & $1.120 \pm 0.039\ [1.081, 1.159]$ &
                       & & & \\
                       & $c$ & $0.616 \pm 0.110\ [0.506, 0.726]$ &
                       & & & \\
\cmidrule{2-7}
Model 2: $a(n+L) + c$  & $a$ & $1.128 \pm 0.035\ [1.093, 1.163]$ &
2.8082 & 0.9795 & \textbf{-285.86} & \textbf{-280.97} \\
                       & $c$ & $0.614 \pm 0.109\ [0.505, 0.723]$ &
                       & & & \\
\bottomrule
\end{tabular}
}
\end{table}

An examination of the empirical results reveals a clear
evolution of the trajectory dynamics depending on the dataset
scale. For the initial 27-point dataset, Model 2 yields the poorest
performance across all metrics, generating the highest RSS ($1.0100$)
and the lowest $R^2_{\text{adj}}$ ($0.9513$). Model 1 shows a
noticeable improvement, extracting a radial slope ($a = 1.275\,\text{GeV}^2$)
that is 16\% larger than the orbital slope
($b = 1.100\,\text{GeV}^2$). Their respective 95\% confidence
intervals show minimal overlap.

As Models~1 and~2 are nested, one can perform a standard nested $F$-test, where
\begin{equation}
\label{Ftest}
    F = \frac{(\text{RSS}_2 - \text{RSS}_1) / (k_1 - k_2)}{\text{RSS}_1 / (N - k_1)}.
\end{equation}
This test yields a $p$-value of 0.0031, thus strongly
rejecting the null hypothesis of linear slope degeneracy, $a=b$,
for the given sample of 27 states. However,
Model~3 turns out to be phenomenologically and statistically superior framework for
these states, achieving the absolute minimum residual variance
($\text{RSS} = 0.6850$) and the lowest information loss
($\text{AIC} = -16.578$, $\text{BIC} = -12.691$). The extracted
positive value of $d = 0.488$ successfully models the empirical
enhancement of mass values of the $L=0$ states.
From the pure statistical viewpoint,
however, the difference between Models~1 and~3 is statistically insignificant since
$\Delta\text{AIC} = \Delta\text{BIC} \simeq 0.43$, which is below the accepted boundary
$\Delta=2$ for inconclusive evidence.

Crucially, a dramatic shift occurs when the sample size is expanded
to the 85-point dataset: The extracted independent
slopes in Model 1 converge, yielding $a = 1.139\,\text{GeV}^2$
and $b = 1.120\,\text{GeV}^2$. The 95\% confidence intervals for these two slopes
now exhibit a substantial and dominant overlap ($(1.096, 1.182)$ versus $(1.081, 1.159)$).

Performing the nested $F$-test~\eqref{Ftest} on the extended dataset yields an $F$-statistic of 0.984 with a
corresponding $p$-value of 0.324 for the null hypothesis that the two slopes are equal.
This $p$-value is significantly greater than the standard 0.05
threshold below which the null hypothesis must be rejected.
Thus, we conclude that the difference between the slope parameters $a$ and $b$ is statistically insignificant.

This convergence of slopes is robustly confirmed by the information criteria. The adjusted coefficient of
determination for both Model~1 and Model~2 becomes completely identical at $0.9795$. As
the  additional parameter in Model~1 fails to deliver any statistically meaningful reduction in
the residual variance, the simpler model is mathematically preferable. Model~2 successfully minimizes the
information metrics, yielding a lower AIC ($-285.86$ versus $-284.87$) and a significantly lower BIC
($-280.97$ versus $-277.55$).  A difference of $\Delta\text{BIC} = 3.42>2$ provides positive statistical
evidence in favor of the universal slope.

Graphically, the orbital Regge trajectories for both models differ little; for Model~2 they are shown in Fig.~1.
The analysis of expanded dataset demonstrates that as higher-lying states are factored into the regression, the relative impact of $L=0$ states
diminishes, and the universal linear trajectory rule predicted by semiclassical string
tension effectively governs the global meson spectrum. To obtain this conclusion, the nonlinear correction is no
longer required, so Model~3 was omitted in Table~\ref{tab:global_comparison} for the extended dataset.
\begin{figure}[htbp]
    \centering
    \includegraphics[width=0.8\textwidth]{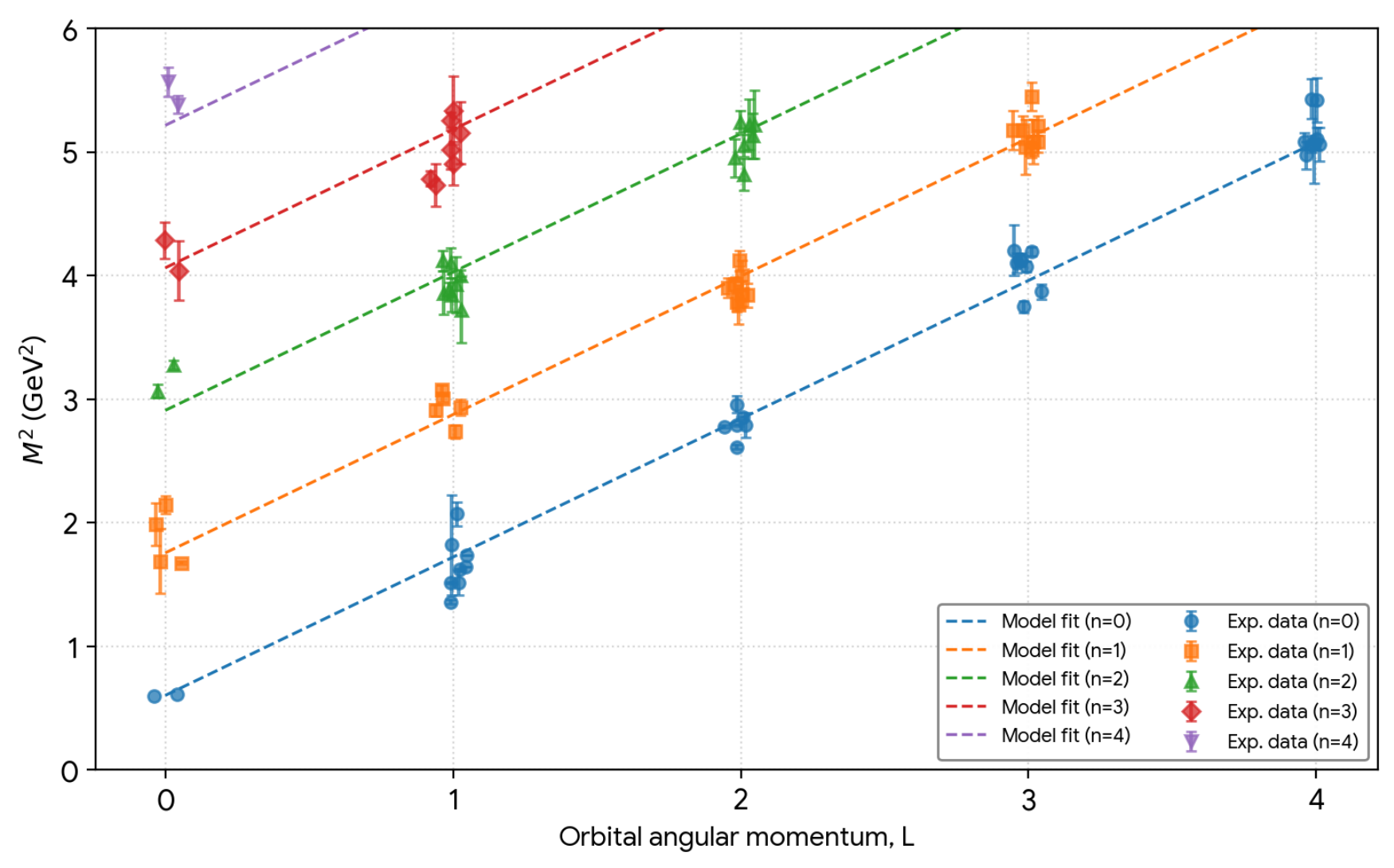}
    \caption{The orbital Regge trajectories for the linear fit with universal slope using all 85 experimental points.}
\end{figure}

\section{Analysis with excluded $S$-wave states}

If we restrict ourselves to the well-known non-strange mesons from PDG, the inequality of slopes of the orbital and radial trajectory would seem obvious. However, a closer look at the experimental masses in Table~2 reveals that the main contribution to this inequality comes from states with $L=0$. This suggests that there are two possibilities for preserving the equality
of slopes: (1) introducing a nonlinear correction in $L$; (2) excluding $S$-wave states from the fit. Model~3 was intentionally introduced above to demonstrate the first possibility
using a specific example. In this section, we demonstrate the second possibility.

Separating out the states with $L=0$ in Table 2 leaves a set of 19 well-established mesons and 73 states in the extended dataset. In the first case, the results are given in Table~\ref{tab:fit1}.
\begin{table}[h!]
\centering
\caption{\small The obtained fit for parameters (in GeV$^2$) in the case of 19-point dataset ($L > 0$) with the corresponding Residual Sum of Squares (RSS)
and Residual Standard Error (RSE).}
\label{tab:fit1}
\vspace{0.3cm}
\begin{tabular}{lcc}
\toprule
\textbf{Parameter / Metric} & \textbf{Model 1}  & \textbf{Model 2} \\
\midrule
$a$ & $1.160 \pm 0.284$ & $1.124 \pm 0.113$ \\
$b$ & $1.122 \pm 0.118$ & — \\
$c$ & $0.543 \pm 0.219$ & $0.544 \pm 0.212$ \\
\midrule
RSS & \textbf{0.4687} & $0.4711$ \\
RSE & $0.1711$ & \textbf{0.1665} \\
AIC & $-10.42$ & \textbf{-12.32} \\
BIC & $-7.59$ & \textbf{-10.44} \\
\bottomrule
\end{tabular}
\end{table}

An $F$-test was conducted to verify whether adding the independent parameter $b$ significantly improves the fit quality.
The test yields an $F$-statistic of $0.0887$, corresponding to a $p$-value of $0.7697$. Since $p > 0.05$, we fail to reject the null hypothesis that $a = b$.
This means that in Model 1, the calculated slopes along the $n$ axis ($a = 1.160$) and the $L$ axis ($b = 1.122$) are statistically indistinguishable within
their respective error margins. Model 2 yields a lower Residual Standard Error ($\text{RSE} = 0.1665$) because eliminating the redundant parameter increases the
degrees of freedom without causing a substantial increase in RSS. Thus, Model~2 is statistically preferable.

The same conclusion follows from comparison of both information criteria --- they are slightly smaller for Model~2. In particular, $\Delta\text{BIC} = 2.85>2$ provides positive statistical evidence in favor of Model~2.

Our analysis extended to an expanded dataset containing 73 experimental points with $L > 0$ (excluding 12 points where $L = 0$) is summarized in Table~\ref{tab:fit2}.

\begin{table}[h!]
\centering
\caption{\small The obtained fit for parameters (in GeV$^2$) in the case of expanded 73-point dataset ($L > 0$).}
\label{tab:fit2}
\vspace{0.3cm}
\begin{tabular}{lcc}
\toprule
\textbf{Parameter / Metric} & \textbf{Model 1} & \textbf{Model 2} \\
\midrule
$a$ & $1.121 \pm 0.045$ & $1.142 \pm 0.039$ \\
$b$ & $1.162 \pm 0.044$ & — \\
$c$ & $0.522 \pm 0.126$ & $0.542 \pm 0.126$ \\
\midrule
RSS & \textbf{1.9322} & $2.0274$ \\
RSE & \textbf{0.1661} & $0.1690$ \\
AIC & \textbf{-51.95} & $-50.45$ \\
BIC & $-45.08$ & \textbf{-45.87} \\
\bottomrule
\end{tabular}
\end{table}

According to the information criteria, the statistical difference between the models here is on the verge of fluctuation.

Comparing the two models for the 73-point sample via $F$-test for nested models yields an $F$-statistic of $3.445$, which results in a $p$-value of $0.0677$.
At the standard 5\% significance level ($\alpha = 0.05$), since $p > 0.05$, the null hypothesis $a = b$ cannot be rejected.
The extracted intervals for the values of radial and orbital slopes in Model~1 ($a = 1.121 \pm 0.045$ and $b = 1.162 \pm 0.044$) exhibit a substantial overlap.
Thus, Model~2 remains statistically superior due to its simplicity, successfully capturing the underlying physical law with fewer parameters.
This strongly validates the hypothesis of a universal slope for trajectories across both quantum numbers $n$ and $L$. Finally, the most economical
and statistically preferable parametrization of spectrum with excluded $S$-wave states is\footnote{Note that the classical Regge trajectory is defined as
$J(t)$, where $J$ is the total angular momentum and the square of transferred momentum $t=M^2$ at a resonance position. The empirical relation~\eqref{cl} can be easily translated into this form. For example, the mesons lying on the leading Regge trajectory have $J=L+1$, then~\eqref{cl} takes the form
\begin{equation}
    J = (0.88\pm0.03)t + 0.53\pm0.11.
\end{equation}
The parameters of this fit are very close to those known for a long time~\cite{Collins:1977}.}
(in GeV$^2$, see Table~\ref{tab:fit2})
\begin{equation}
    M^2 = (1.142\pm0.039)(n+L) + 0.542\pm0.126.
    \label{cl}
\end{equation}

The obtained fit~\eqref{cl} is visualized in Fig.~2.
An examination of the 12 excluded points with $L=0$ reveals a systematic upward shift relative to the main regression line.
This reflects manifestations of short-range effects contributing to masses of the $L=0$ states,
thus justifying exclusion of corresponding resonances from the universal linear fit.
These effects are known to be the most dramatic for the pion. Fig.~2 shows that a tail from these effects persist at higher energies.
\begin{figure}[htbp]
    \centering
    \includegraphics[width=0.7\textwidth]{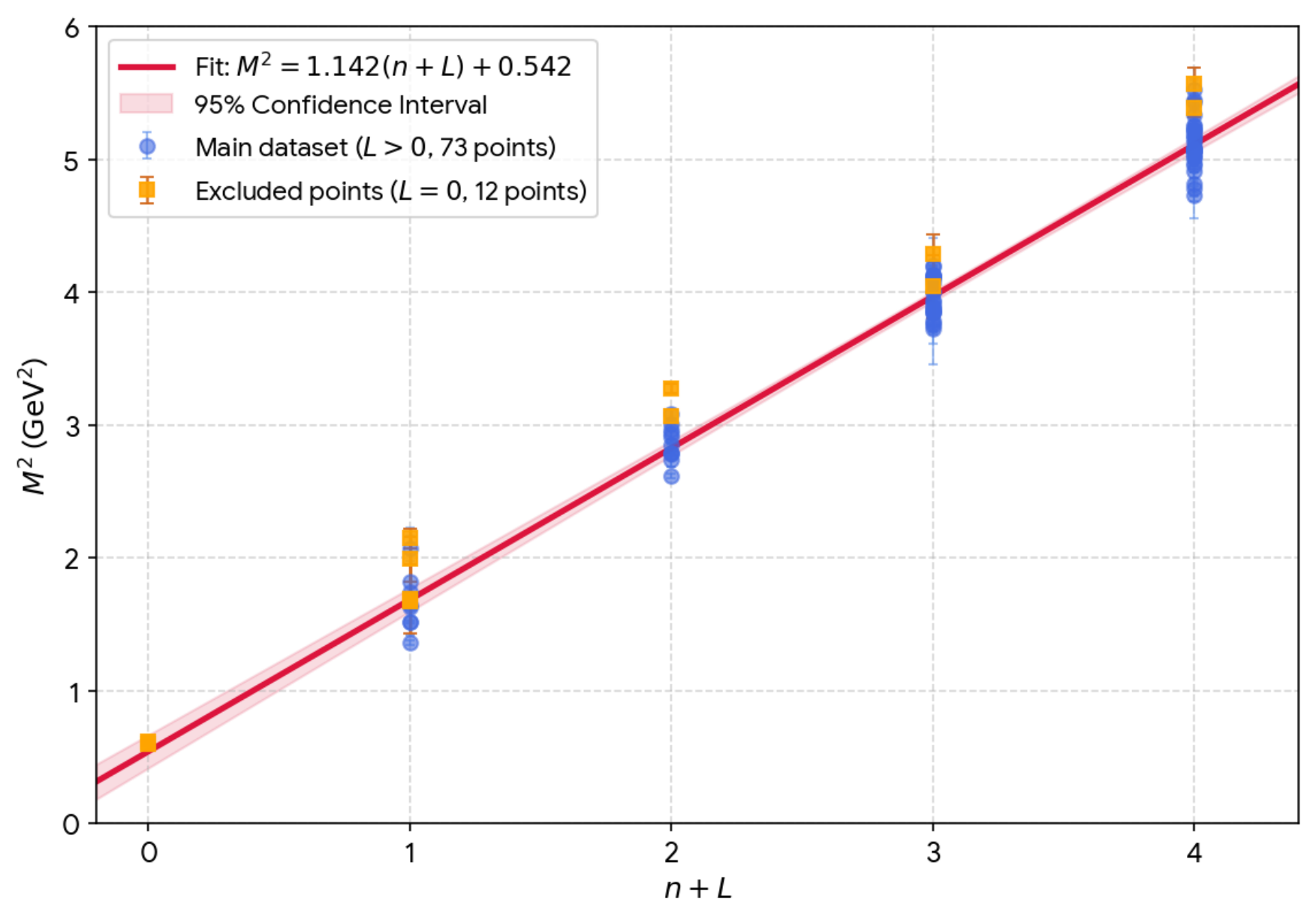}
    \caption{Visualization of fit~\eqref{cl} and the experimental data.}
\end{figure}

Alternatively, the spectrum can be visualized as in Fig.~1 --- as a family of equidistant trajectories  plotted against the orbital angular momentum $L$. They follow with a constant vertical step $\Delta M^2 = a = 1.142\text{ GeV}^2$. The lowest line represents the main trajectory ($n=0$) which is followed by the daughter trajectories ($n=1, 2, 3, 4$). The corresponding plot is given in Fig.~3. The systematic upward shift of masses at $L=0$ is clearly seen.
\begin{figure}[htbp]
    \centering
    \includegraphics[width=0.7\textwidth]{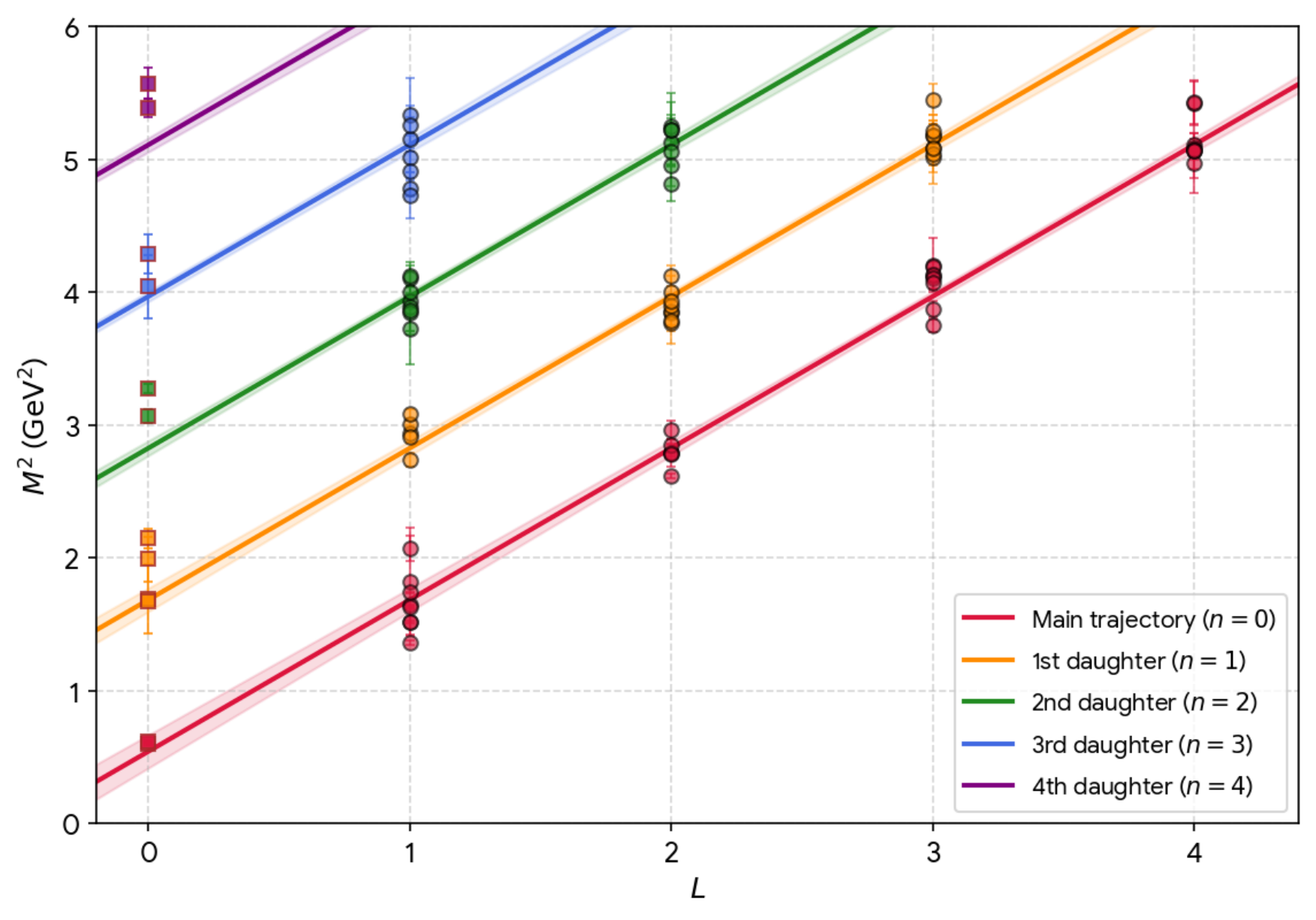}
    \caption{The orbital Regge trajectories following from fit~\eqref{cl}. The excluded $S$-wave states are indicated by squares as in Fig.~2. 95\% confidence intervals are shown.}
\end{figure}

\section{Conclusion}

By employing several information criteria, we performed a statistical analysis of competing Regge-like formulas describing the spectrum of light non-strange mesons --- specifically comparing models with different slopes for orbital and radial trajectories against those with a universal slope. For well-established states from the PDG, the slopes are clearly distinct. We argue, however, that this discrepancy may be an effective difference arising from the omission of non-linear corrections. Such corrections are crucial for accurately describing the spectrum of $S$-wave states. This hypothesis was demonstrated using a simple model with a nonlinear correction of the form $M^2(L,n)\propto a(n+L)+d/(L+1)$. This model possesses the same number of parameters as the variant with different slopes, but proves to be statistically and phenomenologically superior. In other words, the inequality of slopes may simply be an artifact of inappropriately including $L=0$ states into the linear Regge fit. The data indicate that not only the pion but all other $S$-wave mesons are significantly influenced by short-range physics (relative to the confinement scale), which is not captured by a simple semiclassical string model.

In the next step, we extended our analysis using a recently proposed $(L,n)$-classification of light non-strange mesons encompassing 85 states~\cite{Afonin:2025dxk}. Here, the difference between the two Regge models was found to be statistically insignificant. The degenerate linear model, $M^2 = a(n+L) + c$, successfully captures the underlying physical law with fewer degrees of freedom, thereby strongly validating the hypothesis of a universal string tension in the global spectrum of light non-strange mesons. Thus, our results confirm that the spectrum of light non-strange mesons is dominated by Regge behavior with a universal slope for both angular and radial trajectories.

\end{document}